\documentclass[letterpaper, 12pt]{article}%
\usepackage{times}
\usepackage{amsmath}
\usepackage{graphicx}%
\usepackage{authblk}
\usepackage{amsfonts}%
\usepackage{amssymb}
\usepackage{color}
\usepackage{natbib}
\usepackage{bm}
\usepackage{setspace}
\usepackage{lineno}

\newcommand{\bbet}{ \mbox{\boldmath $ \beta $} }
\newcommand{\bbeta}{ \mbox{\boldmath $ \beta $} }

\newcommand{\balpha}{ \mbox{\boldmath $ \alpha $} }

\newcommand{\bs}{\textbf{s}}

\newcommand{\bw}{\textbf{w}}

\newcommand{\bx}{\textbf{x}}

\usepackage{fancyhdr}
\usepackage{lipsum}

\fancypagestyle{titlepage}{
  \lhead{Ecological Monographs, xx(x), 2019, pp. xx-xx}
  \rhead{}

}

\begin{document}

\title{Preferential sampling for presence/absence data and for fusion of presence/absence data with presence-only data}
\author[1]{Alan E. Gelfand}
\author[2]{Shinichiro Shirota}
\affil[1]{Department of Statistical Science, Duke University}
\affil[2]{Department of Biostatistics, UCLA}
\maketitle\thispagestyle{titlepage}

\bigskip
\clearpage
\begin{abstract}

Presence/absence data and presence-only data are the two customary sources for learning about species distributions over a region.  We present an ambitious agenda with regard to the analysis of such data.  We illuminate the fundamental modeling differences between the two types of data. Most simply, locations are considered as fixed under presence/absence data; locations are random under presence-only data.   The definition of ``probability of presence'' is incompatible between the two.  So, we are not comfortable with modeling strategies in the literature which ignore this incompatibility, which assume that presence/absence modeling can be induced from presence-only specifications and therefore, that fusion of presence-only and presence/absence data sources is routine.
While, in some cases, data collection may not support this, we propose that, since, in nature, presence/absence is seen at point locations, presence/absence data should be modeled at point level.  If so, then we need to specify two surfaces.  The first provides the probability of presence at any location in the region. The second provides a realization from this surface in the form of a binary map yielding the results of Bernoulli trials across all locations; this surface is only partially observed.   On the other hand, presence-only data should be modeled as a (partially observed) point pattern, arising from a random number of individuals seen at random locations, driven by specification of an intensity function.  There is no notion of Bernoulli trials; events are associated with areas.
We further suggest that, with just presence/absence data, preferential sampling of locations may arise.  Accounting for this, using a shared process perspective, can improve our estimated presence/absence surface as well as prediction of presence.   We further propose that preferential sampling can enable a probabilistically coherent fusion of the two data types.
We illustrate with two real datasets, one presence/absence, one presence-only, for invasive species presence in New England in the United States.  We demonstrate that potential bias in sampling locations can affect inference with regard to presence/absence and show that inference can be improved with preferential sampling ideas.  We also provide a probabilistically coherent fusion of the two datasets again with the goal of improving inference for presence/absence.
The importance of our work is to encourage more careful modeling when studying species distributions.  Ignoring incompatibility between data types and adopting nongenerative modeling specifications results in invalid inference; the quantitative ecological community should benefit from this recognition.

\end{abstract}

\vspace{.3cm}

\textbf{Keywords: areal unit data; geostatistical model; hierarchical model; logGaussian Cox process; point-referenced data; shared process model}

\section*{Introduction}

Learning about species distributions is an important activity in the ecology community.  The literature discusses two types of data collection to learn about species distributions: presence/absence and presence-only.  The former works with some version of designed sampling where say plots (grid cells, quadrats, etc.) are sampled and presence/absence or abundance of a species is observed for the sampled plots.  That is, locations are fixed.  Presence-only data arises through randomly encountering a species within a region and is typically collected in the form of museum or citizen science data.  That is, locations are random.   In fact, the distinction between the two types of data collection can be murky since, if data collection is developed through gridding of cells, then, conceptually, the observations associated with the cells can be viewed as capturing presence/absence as well as presence-only, as we elaborate below.  In any event, the literature on modeling presence/absence data is enormous by now and, more recently, there has been a consequential growth in the literature addressing modeling for the presence-only setting.  References to this literature are supplied as part of the development in Sections ``Some presence/absence modeling details" and ``Fusing presence/absence and presence-only data", respectively.

The contribution of this paper is to address some fundamental and occasionally contentious threads in the literature with regard to the foregoing data collection.  For instance, it is asserted that a common modeling framework can be used for both data types, that presence/absence data modeling can be induced under a presence-only framework, and, moreover, that presence-only data can be used to infer about presence/absence \citep{Dorazio(14), Royleetal(12), HastieFithian(13)}.  A further implication is that fusion of general presence/absence and presence-only data sources can be implemented within what is essentially the presence-only framework \citep{Pacifici(17)}.

We consider these issues first with discussion to attempt to clarify what ``presence at a location'' means.  We argue that probabilistic modeling for the two data types is distinct and incompatible.  Specifically, since, in nature, presence/absence is seen at point locations, we propose that presence/absence data should be modeled at point level.  We note that, in practice, presence/absence data is not always collected at fine resolution.  Often data collection is such that presence/absence is only recorded as a binary event over an areal unit.  Such data collection loses information by ignoring say, the number of individuals found in the unit and perhaps even information on the locations of the individuals in the unit.  It also suggests that the size of the unit should be considered with regard to the chance of presence in the unit.  Furthermore, this does not refute the assertion that the presence/absence process should be modeled at point-level; rather, it reveals that the data collection makes it difficult to implement point-level modeling.  In practice, the size of the unit is rarely incorporated into the modeling and, in fact, if size is not considered then the modeling returns us to point level specification where the unit is geo-referenced say, by its centroid.  If the size of the unit is incorporated into the modeling then we find ourselves closer to the spirit of presence-only modeling, as we clarify below.

Next, under point-level modeling for such data, we bring in preferential sampling ideas to clarify how potential bias in selection of sampling locations can affect inference with regard to presence/absence.  Using what is referred to as the ``shared process'' perspective, we demonstrate that estimation of the probability of presence as well as prediction of presence can be improved by accounting for preferential sampling.  Then, we briefly turn to the fusion problem, arguing that current versions of such fusion in the literature have fundamental flaws.  We propose a probabilistically coherent fusion, again employing the shared process perspective for implementing the fusion, extending application of preferential sampling.  This allows the two data sources to be probabilistically independent or dependent. Altogether, this perspective provides a collection of models to take presence/absence modeling to a richer explanatory level.

To examine the foregoing issues, we need to look carefully at how presence/absence has been customarily modeled in the literature.  We also need to do the same with regard to the presence-only literature.  Further, we need to elaborate what preferential sampling is in order to reveal its utility for these issues.  In the interest of keeping the explication at a concise and, hopefully, comfortable level, we only consider individual species models.  However, extension to joint species distribution modeling is available and will be presented in a subsequent paper.  In order to go forward, we first offer some preliminary words regarding what a presence/absence event means (to expedite the flow we defer referencing to subsequent sections).

\subsection*{The fundamental issue}
\label{sec:issue}
The fundamental issue that underlies the development of this paper is the attempt to clarify exactly what a presence/absence event means?  It seems that if one asks different ecologists one may get different answers.  As above, for a given species, is it going to a geo-coded location and recording whether or not the species is there?  Or, is it a binary summary for an areal unit?  Was the species observed on the unit or not?  Here, in order to bridge with preferential sampling ideas, we need to conceptualize presence/absence at point-level. In this way, we can formalize a ``probability of presence'' surface which provides the presence/absence probability for the species at each location in the region, driven by environmental features at the location.  Below, we attempt to clarify more about the behavior of this surface.  However, for example, this surface can be thresholded at a selected probability in order to obtain a \emph{niche} for the species within the region. If we add to this surface a conceptual Bernoulli trial at every location, we obtain a realization of a presence/absence surface for the species, a binary map with a $1$ or $0$ at each point in the region. This surface is only partially observed through the data collection.  In terms of ``seeing'' this surface, at best we can display it with a high resolution grid of points.

Therefore, \emph{coherent} modeling for these two surfaces which enables a \emph{generative} probabilistic model for presence/absence data is a primary objective of this paper.  Presence/absence modeling in the context of areal units does not permit modeling of a probability of presence surface; in this case presence/absence probability will depend upon the size, shape, orientation, etc. of the areal units.

We focus on plants (in order to remove movement challenges).  (To attempt to align with animal movement modeling terminology, for plants the term \emph{occupancy} is equivalent to presence and the term \emph{use} might be connected to high probability of presence.)  Then, for a given plant species, we can ask what the \emph{true} realization of a presence/absence surface over a fixed region at a \emph{fixed} time would look like?  At any location in the region, this surface must take on the value of either $1$ if the species is present there or $0$ if it is not.  If this surface is specified to be $1$ for some areal units and $0$ for others, then the realization of the surface will be dependent upon the selection of the units, their size, their shape, their orientation.  This would seem to conflict with how presence/absence arises in nature.  Such incoherence can be avoided if presence/absence is viewed at \emph{point} level.
Furthermore, from a point level definition, we can scale up to arbitrary areal units (see below) whereas we can not do the reverse.

More explicitly, from a point-level perspective, we can ``see'' the realization of the presence/absence surface over the entire region of interest, and thus, over any subset of the region without imposing any areal scales.
In this regard, presence/absence data is frequently associated with areal units, e.g., described as presence/absence over a grid cell, e.g., a plot or a quadrat.  Depending upon the size of the region relative to the size of the areal units, the unit may be considered as a point in the region.  However, formally, presence/absence is \emph{never} observed at a point.  Even at fine resolution, a point is only specified with regard to a number of significant decimal places so, implicitly, it is an area due to rounding.  The idea of a point-level process specification is accepted as conceptual.We adopt this idea routinely in modeling data. We never observe continuous measurements; we only observe them up to decimal accuracy.  Nonetheless, conceptually, we proceed to model them as continuous.

When presence/absence data is recorded at areal units, presence is customarily declared if the species is found anywhere in the unit.  But then, it seems necessary to model the probability of presence considering the size of the unit.   Moreover, this definition would ignore the abundance on the unit.  Should presence associated with one individual in a unit be the same as presence associated with ten individuals in a unit of the same size?  Shouldn't there be implications for probability of presence in the unit?  Coherence finds us wanting to think of presence/absence in a \emph{dimensionless} fashion.

It can be argued that the presences of a species over a region form a point pattern.  That is, there are a random, finite, number of individuals randomly located in the region.  We agree and pursue this line of thinking more precisely below.  However, we seek to make a connection to a realization of a presence/absence surface as well as to a model for a probability of presence/absence surface.  In this regard, would an ecologist who went out to sample a fixed collection of units for presence/absence attempt to model presence/absence through a (partial) realization of a point pattern?  We think, instead, that some version of a regression model using suitable unit-level covariates would be attempted, as we describe below.
Furthermore, we see that there is no notion of an  \emph{intensity} associated with presence/absence observations.  Intensities arise from thinking about presence through point patterns, a perspective that is associated with presence-only data, as we develop below.  Intensity surfaces can be normalized to density surfaces.  Such density surfaces reflect the relative chance of observing a species at a given location compared with other locations in the region.  They have nothing to do with providing a probability of presence surface.

If we scale a realization of a presence/absence surface as a binary map to an areal unit then it makes sense to think about the average of the realization over that unit, i.e., the proportion of $1$'s over the unit.  This proportion is the \emph{empirical} chance of finding the species present at a randomly selected location within the unit.  In fact, the proportion  of $1$'s over the entire region can be interpreted as the prevalence of the species over the region.  Similarly, with a modeled probability of presence surface, if we scale this surface over the unit, we obtain an average probability over the unit.  This average conveys the modeled probability of finding a presence at a randomly selected location within the unit.  The issue is that we need not think in terms of areal units in order to model presence/absence.  If we want to investigate units then we can scale accordingly.

\section*{Our motivating dataset}\label{sec:Data}

We illustrate all of the above with invasive plant data from New England in the U.S.
We extracted a subregion of the six New England states (Connecticut, Rhode Island, Massachusetts, Vermont, New Hampshire and Maine).  The presence/absence dataset comes from the Invasive Plant Atlas of New England (IPANE) and consists of more than $4000$ sites where invasive species surveys were conducted and focuses on seven species.  Details are provided below.  The presence-only dataset comes from the Global Biodiversity Information Facility (GBIF) which is a data aggregator for biological collections worldwide.  The number of observations will vary from species to species.  Details are provided below.

IPANE is a citizen science organization that engages volunteers in scientifically rigorous sampling protocols.
There are 4314 unique sampling sites across New England where invasive plant surveys were conducted.
Each site is provided with a location (latitude, longitude) and has been classified with regard to each focal species as a presence (focal species recorded) or an absence (focal species not recorded).
The dataset includes seven of the most common invasive plant species in the IPANE database: multiflora rose (MR), oriental bittersweet (OB), Japanese barberry (JB), glossy buckthorn (GB), autumn olive (AO), burning bush (BB) and garlic mustard (GM).
All species are terrestrial and all but garlic mustard are woody (shrubs, small trees, or vines).
These species vary in their land cover associations (e.g., some occur in forest understory and others occur in open habitats).
We consider the same species within GBIF.
Duplicated points and points lying outside the study region are discarded from the original dataset.
Table \ref{tab:name} displays the species name and sample size for the IPANE and the GBIF datasets.
In the analysis below, for convenience, longitude and latitude are transformed to eastings and northings, and rescaled from km units to 10 km units.

\begin{figure}[ht]
  \begin{center}
   \includegraphics[width=15cm]{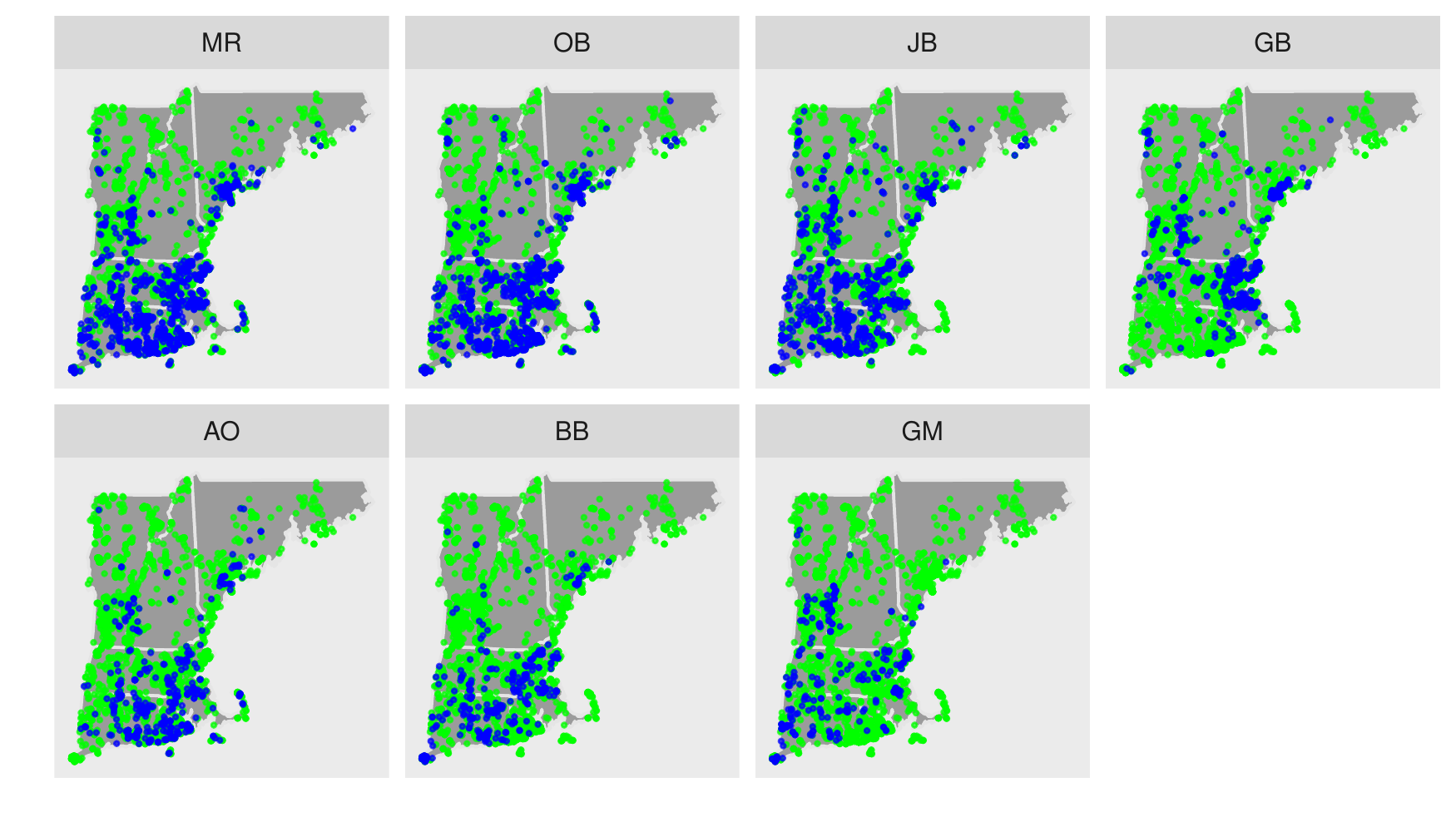}
  \end{center}
  \caption{}
  \label{fig:PAlocations}
\end{figure}

\begin{figure}[ht]
  \begin{center}
   \includegraphics[width=15cm]{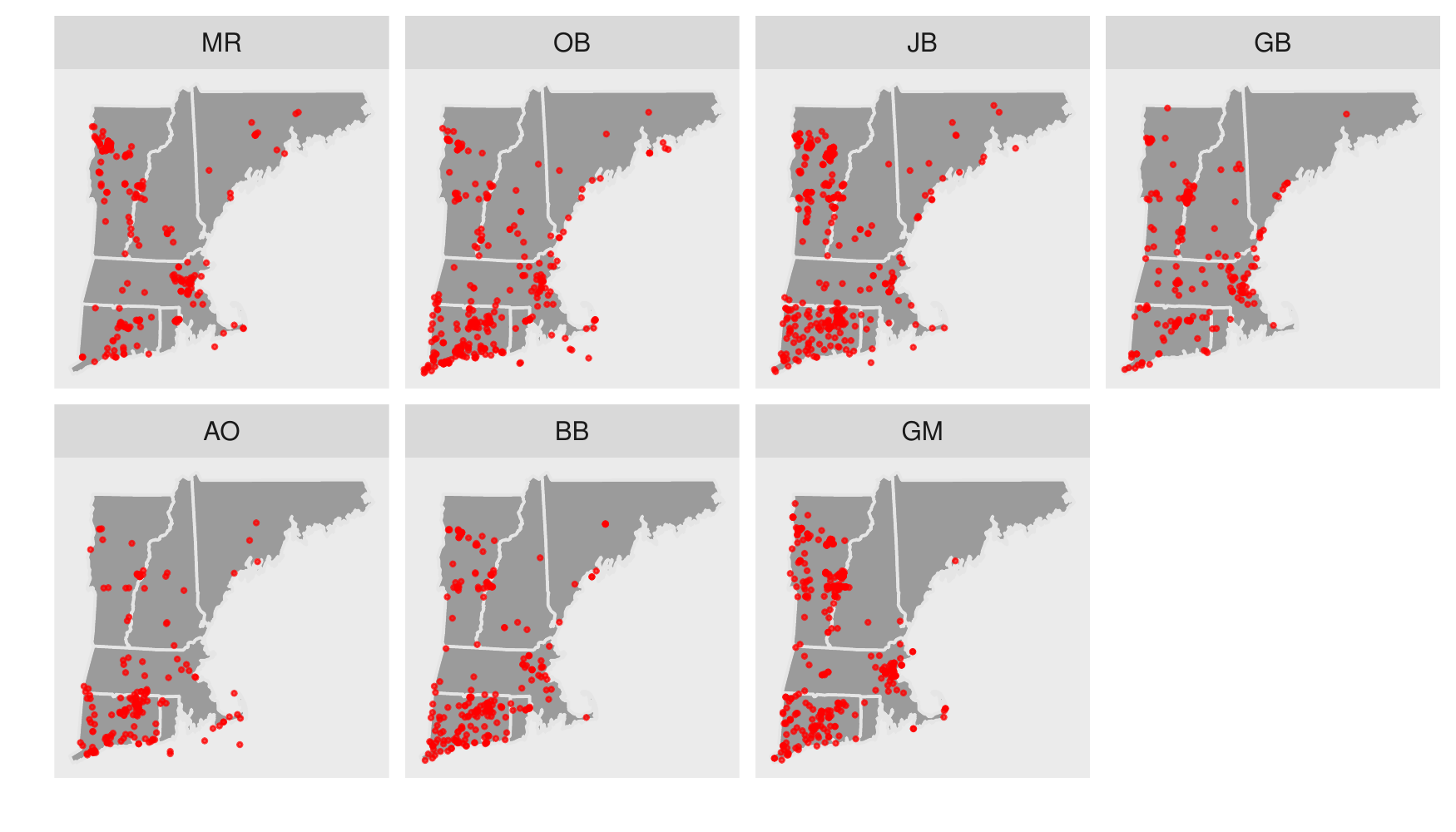}
  \end{center}
  \caption{}
  \label{fig:POlocations}
\end{figure}

Figure \ref{fig:PAlocations} displays the distribution of presence and absence locations from IPANE for each species across the study region.
Figure \ref{fig:POlocations} displays the distribution of presence-only points from GBIF for each species across the study region.
For some species, for example garlic mustard, the distribution of the presence-only points shows a different pattern from that of the observed presences in the presence/absence data.  Once more, the presences in IPANE arise from fixed sampling locations while the presences in GBIF arise at random locations.  Importantly, we removed all of upper Maine as the figures show.  Both the IPANE and the GBIF data were so sparse there that extending spatial modeling to include this region produced poorly behaved model fitting.

\begin{figure}[ht]
  \begin{center}
   \includegraphics[width=15cm]{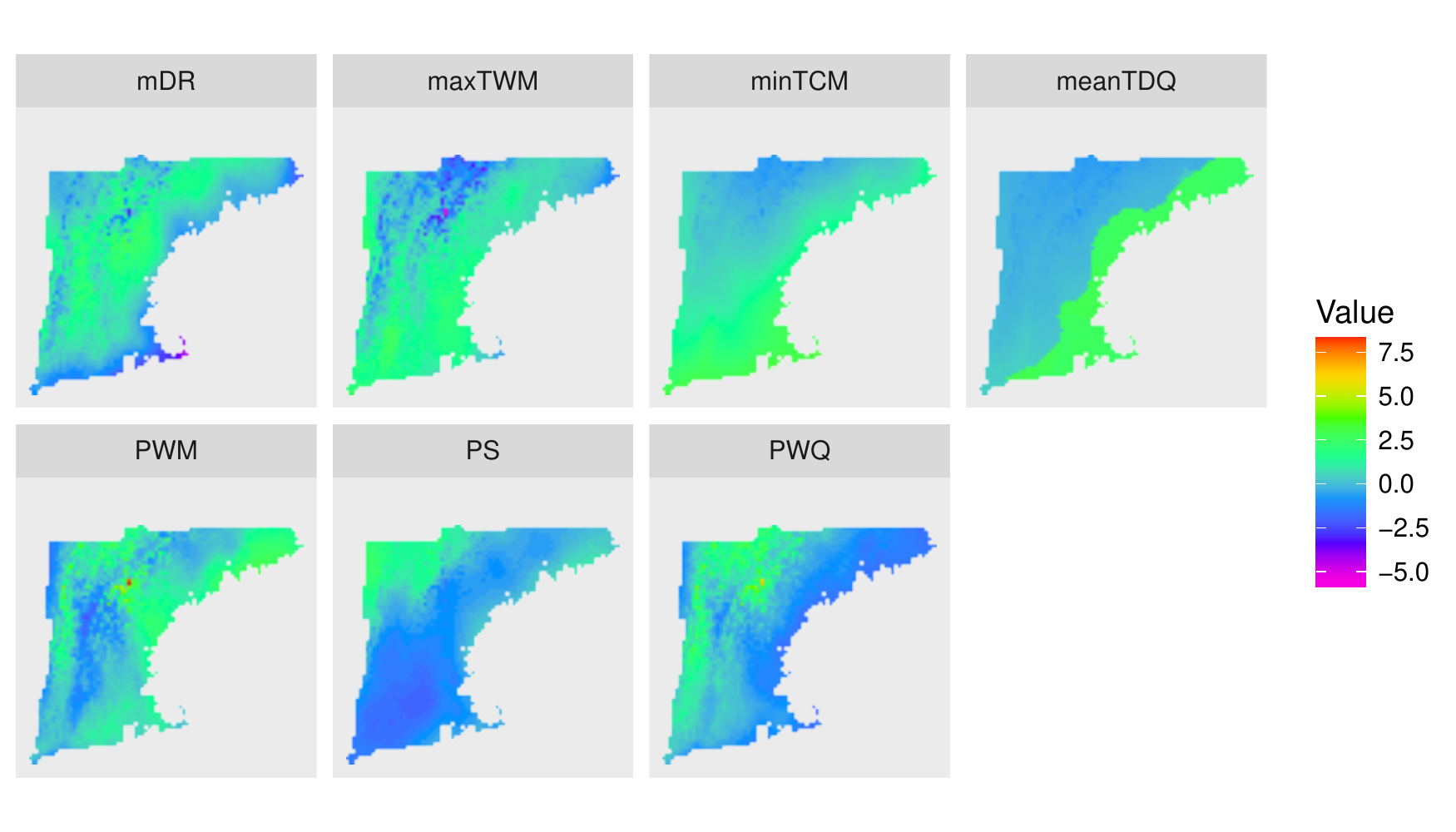}
  \end{center}
  \caption{}
  \label{fig:Covsurface}
\end{figure}

Adding to the original database, we have 19 potential covariates provided by WorldClim (version 1.4, http:/ /www.worldclim.org/version1) as 30-arc second ($\sim$1 km) raster data. We select 7 covariates from them by discarding highly correlated covariates.  They are (1) mean diurnal range (mDR, mean of monthly (max temp-min temp)), (2) max temperature of warmest month (maxTWM), (3) min temperature of coldest month (minTCM), (4) mean temperature of driest quarter (meanTDQ), (5) precipitation of wettest month (PWM), (6) precipitation seasonality (PS, the standard deviation of the monthly precipitation estimates expressed as a percentage of the mean of those estimates, that is, the annual mean), and (7) precipitation of warmest quarter (PWQ).   With regard to possible multicollinearity concerns, these seven covariates were chosen such that each pair has absolute correlation less than 0.7.
Figure \ref{fig:Covsurface} displays the standardized covariate surfaces for the 7 selected covariates.
The location that indicates extreme values in maxTWM and PWM corresponds to the summit of Mt. Washington which is notorious for exhibiting extreme climate conditions.



\section*{Some presence/absence modeling details}
\label{sec:P/A}
Confining ourselves to plants and following the discussion in the Introduction, we make the assumption that presence/absence data arises as observation of binary responses, presence (1) or absence (0) at a collection of sampling locations \citep[see, e.g.,][and references therein for a review]{Elithetal(06)}.  The goal is to explain the probability of presence at a location given the environmental conditions that are present there.  The customary approach is to build a binary regression model with say a logit or probit link where the covariates can be introduced linearly (see below) or as smoothly varying functions.  The latter choice results in generalized additive models (GAMs) which tend to fit data well since they employ additional parameters to enable the response to assume nonlinear and multimodal relationships with the predictors \citep[][]{Guisanetal(02), Elithetal(06)}.  The price that is paid for using GAMs is a loss of simplicity in interpretation as well as the risk of overfitting resulting in poor out-of-sample prediction.  We don't consider GAMs further here.

Much of the early presence/absence work was \emph{non-spatial} in the sense that, in modeling presence/absence probabilities, though spatial covariate information was included, potential spatial dependence in the residuals was not.  Accounting for the latter seems critical. Causal ecological explanations such as localized dispersal as well as omitted (or unobserved) explanatory variables with spatial pattern such as local smoothness of soil or topographic features suggest that, at sufficiently high resolution, occurrence of a species at one location will be associated with its occurrence at neighboring locations \citep[][]{VerHoefetal(01)}. In particular, such dependence structure, introduced through spatial random effects, facilitates learning about presence/absence for portions of a study region that have not been sampled, accommodating gaps in sampling and irregular sampling effort.

Following the framework presented in \cite{Gelfandetal(06)}, suppose $Y(\bs)$ denotes the presence/absence (1/0) of the
species at sample location $\bs$. If the study region $D$ is partitioned into grid cells, say at the level of resolution of the environmental covariates, then, summing up $Y(\bs)$ over $n_i$, the
number of sites sampled in cell $i$, yields grid cell level
counts: $Y_{i+}=\sum_{\bs\in \text{grid} i}Y(\bs)$. This is an elementary illustration of scaling up from points to areal units. If the sampling site is viewed as the grid cell then we have $n_{i}=1$, a single Bernoulli trial for the cell.  If the cell was not sampled, we have $n_{i} = 0$.

If we assume
independence for the trials, a binomial distribution results for
$Y_{i+}$, i.e., $Y_{i+} \sim Binomial(n_i, p_i)$.
Explicitly, the probability that the species occurs in cell
$i$, $p_i$, is related functionally to the environmental
variables with a logit link function and a linear (in coefficients) predictor ${\bw}^{T}_i \bbet$, e.g.,
$\log \left(\frac{p_i}{1-p_i}\right)={\bw}^{T}_i \bbet $.
Here ${\bw}_i$ is a vector of explanatory environmental variables associated with cell $i$ and $\bbet$ is a vector of
associated coefficients.  Here, and in the sequel, we could equally well use a probit link function.

We can extend this grid cell level model to be spatially explicit
by adding spatial random effects.  In modeling $p_i$, a spatial term $\rho_i$
associated with grid $i$ is added yielding $ \text{log}\left(\frac{p_i}{1-p_i}\right)=\bw^{T}_i \bbet + \rho_i$.
The random effect $\rho_i$ adjusts the probability of presence of the modeled
species up or down, depending on the values in a \emph{spatial neighborhood} of cell $i$. To capture this behavior, we customarily employ a Gaussian intrinsic or conditional auto-regressive (CAR) model \citep{Besag(74)}.  Such a model proposes that the effect for a
particular grid cell should be roughly the average of the effects of its neighboring cells and results in a multivariate
normal as the joint distribution over all the cells. There are many ways to specify neighbor structure; see \cite{BanerjeeCarlinGelfand(14)} for a full discussion.

%

If we view the visited sites as points and therefore model at point scale, $Y(\bs)$ would be taken as
\begin{equation}
Y(\bs) \sim Bernoulli(p(\bs)), \label{distributionofysite}
\end{equation}
analogously relating the probability that the species occurs in site $\bs$, $p(\bs)$, to the set of environmental
variables as $\log \left(\frac{p(\bs)}{1-p(\bs)}\right)={\bw^{T}(\bs)} \bbet$. Such modelling requires that we have covariate levels $\bw(\bs)$
for each site. This model is referred to as a spatial regression in the sense that the regressors are spatially referenced.  If we set $\bw(\bs)=\bw_i$ when $\bs$ is within grid $i$, we return to the grid cell model above.

Most relevant for the remainder of this paper, we extend (1) to bring in spatial dependence between points based on their relative locations using Gaussian processes, creating geostatistical models \citep{BanerjeeCarlinGelfand(14)}.  We would model $Y(\bs)$ given $p(\bs)$ and augment the
explanation of $p(\bs)$ through the form
\begin{equation}
\log \frac{p(\bs)}{1-p(\bs)}={\bw^{T}(\bs)} \bbet +\omega(\bs).
\label{logitreg}
\end{equation}
Here,
$\omega(\bs)$ is the spatial random effect associated with point $\bs$, arising as a realization of a Gaussian process.  A suitable covariance function would be selected.
With binary response, this model is referred to a spatial generalized linear model (GLM); see \cite{DiggleTawnMoyeed(98)}.
The model has two levels: the first or data-level specification is a Bernoulli trial and the second or process level presents the probability of presence surface. Inference from \eqref{logitreg} would be about this surface at any location in the study region, with these probabilities explained through the spatially referenced predictors.
The realized presence/absence surface, i.e. $\{Y(\bs): \bs \in D\}$ associated with the first level is also of interest.

The fact that, in practice, presence/absence is not observable at point level does not preclude useful point level modeling. Indeed, this is the case with all geostatistical modeling \citep{BanerjeeCarlinGelfand(14)}, e.g., temperature is never observed at a dimensionless location but we routinely model temperature surfaces.
Assuming data of the form, $(\bw(\bs_{i}), Y(\bs_{i}))$ for sites $i=1,2,...,n$ and adopting the hierarchical (multi-level) regression in \eqref{logitreg} with say, a probit link, $P(Y(\bs)=1) \equiv p(\bs) = \Phi(\bw^{T}(\bs)\bbeta + \omega(\bs))$.  That is, $P(Y(\bs)=1) = P(Z(\bs) >0)$ where $Z(\bs) = \bw^{T}(\bs)\bbeta + \omega(\bs) + \epsilon(\bs)$. Here, $\epsilon(\bs)$ is pure error, i.e., $\epsilon \sim N(0,1)$ and $\omega(\bs)$ is a mean $0$ Gaussian process with a suitable correlation function, typically an exponential or, more generally, a Mat\'{e}rn.  See, e.g., \cite{BanerjeeCarlinGelfand(14)} Chapter 6 for full discussion of such regression models.

Under this model, the $Y(\bs)$ are drawn as conditionally independent Bernoulli trials given $p(\bs)$ (and the associated $Z(\bs)$ are conditionally independent normals).  As a result, even if the probability of presence surface $p(\bs)$ is smooth,  realizations of the presence/absence surface are everywhere discontinuous. Below (``What does ``probability of presence'' mean?") we suggest that, under a point-level modeling specification, such behavior may not be desirable.  In the Supplementary Material (Appendix S2) we present an alternative specification which deals with this concern as well as an associated technical problem.

\section*{What does ``probability of presence'' mean?}
\label{sec:ppmean}

Following ``Introduction: The fundamental issue'' we now attempt more explicit discussion regarding what an observed presence means and the associated implications.
The issue is whether presence/absence is viewed as an event at point level or at areal level.  Is it a Bernoulli trial at say location $\bs$ or is it the probability that the number of individuals of a species in a set, say $A$, is $\geq 1$?

If we model presence/absence at point level, then $Y(\bs)=1$ is a Bernoulli trial at location $\bs$.  However, what does $Y(A)$ mean?  A coherent probabilistic definition specifies it as a block average, i.e., a realization of $Y(A)$ is $Y(A) = \int_{A} 1(Y(\bs)=1)d\bs/|A|$ (where $|A|$ is the area of $A$).  It is the proportion of the $Y(\bs)$ in $A$ that equal $1$; it is not a Bernoulli trial and $P(Y(A)=1) = 0$ since the probability that almost every Bernoulli trial in $A$ results in a $1$ equals $0$.  We can calculate $E(Y(A)) = \int_{A} p(\bs)d\bs/|A|$  with $p(\bs)$ as in \eqref{logitreg}.  That is, $E(Y(A))$ becomes the average probability of presence over $A$.  It is the probability that, at a randomly selected location in $A$, the species is present. If $p(\bs)$ is constant over $A$ then $E(Y(A))$ is this constant probability.
It is interpreted at point level; it is the probability of presence at any site in $A$.

Now, suppose we consider the locations of all individuals in a study region as a random point pattern.  Then, if $N(A)$ is the number of individuals in set $A$, $P(\text{presence} \hspace{.2cm} in \hspace{.2cm} A) = P(N(A) \geq 1)$.  Here, assuming a nonhomogeneous Poisson process (NHPP) or, more generally a log Gaussian Cox process (LGCP) with intensity $\lambda(\bs)$ (see Illian et al. (2008) for a full discussion of NHPPs and LGCPs), $N(A) \sim Po(\lambda(A)) $ where $\lambda(A) = \int_{A} \lambda(s)ds$.  Then, taking the areal unit definition of a presence in $A$, we seek $P(Y(A)=1) = P(N(A) \geq 1) = 1 - e ^{-\lambda(A)}$.
Since presence-only data samples the point pattern (although likely not fully but, rather, up to sampling effort over the region \citep[][]{Chakrabortyetal(11), Fithianetal(15)}), it is compatible with this definition of presence/absence.  However, the probability of a presence in $A$ is only defined with regard to the size of $A$ and will vary with $A$, a concern raised in \cite{HastieFithian(13)}.  As a result, it is unclear how to specify a meaningful probability of presence surface.  Furthermore, the definition of probability of presence as ``one or more'' observations of the species in $A$ yields local distortion to any such surface; $N(A)=1$ or $N(A)=11$ are treated the same with regard to probability of presence in $A$ \citep[][]{Aartsetal(12)}.

The two foregoing definitions associated with $P(\text{presence} \hspace{.2cm} in \hspace{.2cm} A)$ are incompatible and the fundamental difference between them seems to have been missed in the literature.  The conceptualization for the first choice is that we go to fixed ``point'' locations and see what is there; we are not sampling a point pattern.  We model a surface over a domain $D$ which captures the probability of presence at every location in $D$.  The conceptualization for the second is that we identify an area of interest $D$ and, conceptually, we census it completely for all of the occurrences of the point pattern.  We model an intensity which, using the definition above, provides a probability of presence for $A$.
The intensity surface can be normalized to a density surface 
under which the probability of an event at a dimensionless point is 0. This has nothing to do with modeling a Bernoulli trial at a point by specifying a probability of presence at the point, hence a probability of presence surface, through say a probit or logistic regression.

Furthermore, if presented with a collection of plots and observed presence/absence for those plots, one would not model the data as a point pattern.  No point pattern was observed; there is no way to model an intensity.  We would use one of the foregoing presence/absence regression models. Extending to the data fusion problem, suppose one obtains an additional dataset of presence-only observations for the region.  While we could try to model this new dataset as a point pattern, why would it be appropriate to now model the original presence/absence data using a point pattern model associated with the presence-only data?

So, we have articulated the need for care in terms of formalizing the notion of presence of a species as well as the challenge of fusing presence/absence and presence-only data.
In the literature to date, ignoring the incompatibility associated with the scaling issue is the way that presence-only data has been used to provide presence/absence probabilities and also the way presence-only data has been fused with presence/absence data (see, e.g., \cite{Pacifici(17)}).
Instead, we propose probabilistically coherent remediation for this incompatibility in Sections ``Preferential sampling'' and ``Fusing presence/absence and presence-only data'' below.  However, first, in the next subsection, we attempt further clarification of point-level presence/absence modeling.

\subsection*{Further clarification of point level presence/absence modeling}

In reconciling the differences above it may be useful to think more carefully about what the distribution of a species looks like within a specified region, $D$.  Suppose we consider the complete census of individuals in the region.  To be realistic, we have to view the number of presences in a bounded region as finite and therefore a presence must be bigger than a (dimensionless) point since there are an uncountable number of points in $D$.  The scaling issue arises once more.  Formally, a presence can not arise at a point, it is not dimensionless in size; practically, it can be \emph{observed} as point-referenced.  So, at point-level, the presence/absence surface over the region consists of a finite set of ``patches'' where the species is present and, outside of these patches, the species is absent.  From an ecological and practical perspective, we could think of a patch as a collection of individuals of a particular species (it might be just one) that is dense enough so that, at point level, we would declare presence for every location in the patch.  However, if the gaps between the individuals become sufficiently large, then those locations in the gaps must now become absences.  The scaling here is qualitative, not quantitative - an ecologist would not attempt to be precise here and the denseness needed to define a patch depends upon the sizes of the patch relative to the size of $D$.  In the sequel, we also avoid defining patch sizes.

Then, a presence-only realization over the region becomes this finite set of patches.  To view it as a point pattern, we might assume that each individual is \emph{located} at the centroid of its patch.    That is, with a complete census, the number of patches equals the number of points in the point pattern.  However, with regard to a dimensionless definition of presence/absence, a presence at a location is observed if the location falls within a patch associated with a point. This definition of the realized presence/absence surface gives an immediately rigorous definition of prevalence. The prevalence of the species over $D$ is the total area of the patches for the species relative to the total area of $D$.

Some implications are as follows.  First, the number of presence points in $D$ is uncountable, as is the number of absence points.  Second, presence/absence is a \emph{neighborhood} phenomenon.  If there is a presence at $\bs$ then there is presence everywhere in a sufficiently small neighborhood, $\partial \bs$, of $\bs$.  Similarly, if there is an absence at $\bs$, then there must be a neighborhood of $\bs$ where every location is an absence.  As a result, the \emph{realized} presence absence surface is locally constant, i.e., it takes the value $1$ in a patch and $0$ if not in a patch.  A suitable probability model for presence/absence should provide realizations which are locally constant.  This returns us to the discussion at the end of ``Some presence/absence modeling details'' and in Appendix S2 of the Supplementary Material.  A model which assumes conditionally independent Bernoulli trials across locations is not formally appropriate since such a model will provide random $0$s and $1$s across locations, yielding no local constancy.  However, in practice, a suitable version of such a model will usually perform well (see Appendix S2 of the Supplementary Material) and, in fact, such a model is adopted below for computational convenience.

Third, conceptually, the number points in the point pattern can be smaller or larger than the number of observed presence locations.  That is, observing a presence at a location is not identical to observing the centroid associated with the patch containing the observed presence. According to selection of sampling sites, the same individual may be observed at more than one point (though, in practice, it is not likely to be recorded as such) but also, some individuals may never be observed.  Practically, we acknowledge that presence/absence sampling will never observe all individuals but also, that presence-only sampling will rarely observe all individuals. So, formally, with a dataset of point-level presence/absence locations and a dataset of presence-only random locations, at sufficiently fine spatial resolution, the two sets of locations will be disjoint.


\section*{Preferential sampling}
\label{sec:pref}

Working in a point-referenced framework, we bring in preferential sampling ideas to clarify how potential bias in selection of sampling locations can affect inference with regard to presence/absence. Consideration of preferential sampling can improve presence/absence prediction as well as providing modeling for fusing presence-only data with presence/absence data.

\subsection*{What is preferential sampling all about?}

\begin{figure}[ht]
  \begin{center}
   \includegraphics[width=15cm]{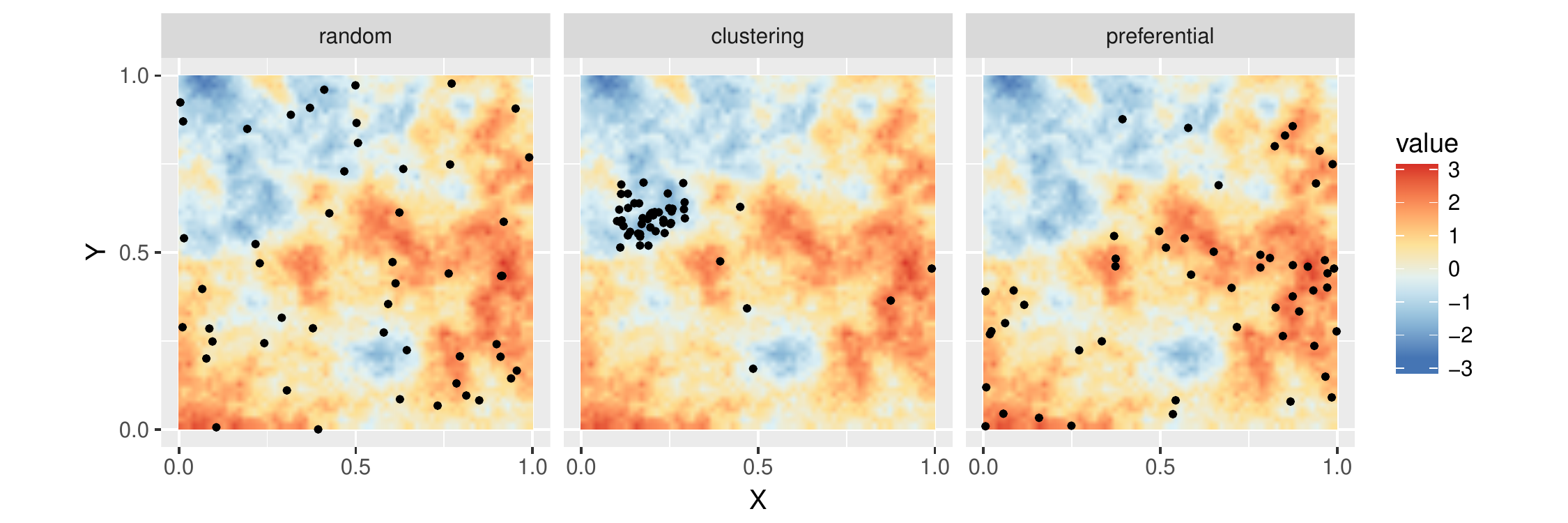}
  \end{center}
  \caption{}
  \label{fig:pref}
\end{figure}

The notion of preferential sampling was introduced into the literature in the seminal paper of \cite{Diggleetal(10)}.  Subsequently, there has been considerable follow up research.  Two useful papers in this regard are \cite{Patietal(11)} and \cite{Cecconietal(16)}. A standard illustration arises in geostatistical modeling \citep[see e.g.][]{CressieWikle(11), BanerjeeCarlinGelfand(14)}.  Consider the objective of inferring about environmental exposures.  If environmental monitors are only placed in locations where environmental levels tend to be high, then interpolation based upon observations from these stations will necessarily produce only high predictions.  A remedy lies in suitable spatial design of the locations, e.g., a random or space-filling design \citep[][]{SaltzmanNychka(98)} for locations over the region of interest is expected to preclude such bias.  Figure \ref{fig:pref} presents three examples showing choices of presence/absence sampling locations relative to a (simulated) response surface.  Interpolation under the preferential sampling scenario will tend to produce predictions which are too high in the blue regions.

In practice, sampling for presence/absence may be designed such that ecologists will tend to sample where they expect to find individuals.  Such \emph{bias} in the collection of sampling locations can affect predictive performance.  Recognizing the possibility of such bias, can we revise presence/absence prediction to adjust for it?  This is the intention of preferential sampling modeling.

While the set of sampling locations may not have been developed randomly, we study it as if it was a realization of a spatial point process. That is, it may be designed/specified in some fashion but not necessarily with the intention of being roughly uniformly distributed over $D$.  Then, the question becomes a stochastic one:  is the realization of the responses independent of the realization of the locations?  If no, then we have what is called preferential sampling.  Importantly, the dependence here is stochastic dependence.  Notationally/functionally, the responses are associated with the locations.  We make this more clear below.


In our context, as we have discussed, the presence/absence data is driven by a probability of presence surface.  This surface plays the role of the ``exposure'' surface, with the observed set of binary responses, ${\cal Y} = (Y(\bs_{1}), Y(\bs_{2}),..., Y(\bs_{n}))$, informing about it. Taking the set of sampling locations as a realization of a random point pattern, ${\cal S} = \{\bs_{1}, \bs_{2},..., \bs_{n}\}$, the question we ask is whether ${\cal Y}$ is independent of ${\cal S}$, again in a stochastic sense?  The answer will depend on the models we supply for ${\cal Y}$ and ${\cal S}$.  Below, we develop several choices, using the idea of a \emph{shared} process, that enable us to address this question and, furthermore, whether ${\cal S}$ enables us to improve our inference regarding prediction of presence for a species at a location.

\subsection*{Preferential sampling models for presence/absence data}
\label{sec:prefP/A}

To develop the stochastic specifications that formalize preferential sampling for a region $D$, we consider two cases for the intensity associated with the point pattern of sampling locations, ${\cal S}$:\\

(i)  $\log \lambda(\bs) = \bw^{T}(\bs)\bbeta$, i.e., a  nonhomogeneous Poisson process (NHPP) and\\

(ii) $\log \lambda(\bs) = \bw^{T}(\bs)\bbeta + \eta(\bs)$, a logGaussian Cox process (LGCP).\\

\noindent  Here, $\bw(\bs)$ is a vector of predictors with associated regression coefficients $\bbeta$ and $\eta(\bs)$ is a mean $0$ GP (below, for convenience, with an exponential covariance function).  See, e.g., \cite{Illianetal(08)} for full discussion of NHPPs and LGCPs. In the sequel we only work with (ii).


We adopt a \emph{direct} model for $Y(\bs)$ through a latent Gaussian process, $Z(\bs)$, i.e., $Y(\bs) = 1(Z(\bs)>0)$, as in the Supplementary Material (Appendix S2).   So, we only need to propose models for $Z(\bs)$.
We start with a simple spatial regression,\\

(a) $Z(\bs) = \bx^{T}(\bs)\balpha + \epsilon(\bs)$,\\

\noindent where the predictors in $\bx(\bs)$ and those in $\bw(\bs)$ need not be identical and $\epsilon(\bs)$ is a pure error with homogeneous variance.  Extension to a customary geostatistical model for $Z(\bs)$ \citep{BanerjeeCarlinGelfand(14)} becomes\\

(b) $Z(\bs) = \bx^{T}(\bs)\balpha + \omega(\bs) + \epsilon(\bs)$,\\

\noindent adding $\omega(\bs)$ as a mean $0$ GP (also with an exponential covariance function for convenience), independent of $\eta(\bs)$ above.  So, using (ii) to model ${\cal S}$ with (b) to model $Y(\bs)$ through $Z(\bs)$, ${\cal S}$ and ${\cal Y}$ are probabilistically independent; there is no preferential sampling.


\cite{Patietal(11)}  attempt to interpret $\eta(\bs)$ as a regressor to add to the geostatistical model for ${\cal Y}$.  That is, they extend (b) to model\\

(c): $Z(\bs) = \bx^{T}(\bs)\balpha + \delta \eta(\bs) +  \omega(\bs) + \epsilon(\bs)$.\\

\noindent Here, the coefficient $\delta$ plays a preferential sampling role.  For example, suppose the design ${\cal S}$ over-samples locations in $D$ where we observe presences, where $Y(\bs)$ tends to be $1$, i.e., where $Z(\bs)$ tends to be high.  Then, $\eta(\bs)$ will tend to be high around those locations.  Therefore, $\eta(\bs)$ can be a significant predictor for $Z(\bs)$ (hence for $Y(\bs)$) with $\delta >0$.  (A similar argument applies when $\delta <0$.) With (ii) and (c), $\eta(\bs)$ is the shared process.

A further shared process model for ${\cal Y}$ that can be explored in this regard extends (a) to \\

(d): $Z(\bs) = \bx^{T}(\bs)\balpha + \delta \eta(\bs) + \epsilon(\bs)$.\\

\noindent Here, interest is in comparing (d) and (ii) with (a) and (ii); is $\delta \neq 0$, i.e., we have a shared process model?
\cite{Diggleetal(10)} focus on comparing (b) and (i) with (b) and (iii). \cite{Patietal(11)} focus on comparing (ii) and (b) with (ii) and (c).

Further modeling possibilities are considered in the Supplementary Material Appendix S3 and the associated Table 2.  However, in the next subsection, we examine just a subset of possible model comparison.  We compare (a) and (ii) with (d) and (ii).
We compare (b) and (ii) with (c) and (ii).
Since the intent is to improve the predictive performance of the model for ${\cal Y}$, model comparison criteria focuses on out-of-sample prediction for $Y(\bs)$'s.

\subsection*{Model fitting and inference for presence/absence data using preferential sampling}
\label{sec:fitting-preferential}

To make the model comparison between (b) with (ii) vs. (c) with (ii) we only need to fit the latter and look at the posterior distribution for $\delta$.
Similarly, for the model comparison between (a) with (ii) vs. (d) with (ii) we only need to fit the latter.  
We do this below for each of the seven species.

We fit models (a) - (d) for the presence/absence data.
For model (c) and (d), we include log Gaussian Cox process models for $\mathcal{S}$, i.e., for model (ii), by taking 2,666 regular grid cells over $D$ to approximate the likelihood over the region.
The regular grid is needed because we introduce the $\bm{\eta}$ surface into models (c) and (d).
Among these grid cells, 1870 do not include any presence/absence locations.
For all species, we use the same seven covariates presented in ``Our motivating dataset'' for both $\bw$ and $\bx$.

These hierarchical models are fitted within a Bayesian framework.  Model fitting details are given in the Supplementary Material (Appendix S4).  As for Bayesian inference, although Gibbs sampling is available for the $\omega(\bs)$ process, its computational cost/time is $\mathcal{O}(n^3)$ and required memory is $\mathcal{O}(n^2)$.
In our case, we have a relatively large $n=4314$, so we implement a nearest neighbor Gaussian process \citep[][NNGP]{Dattaetal(16a)}, which is a sparse Gaussian process model whose computational time is $\mathcal{O}(nk^3)$ (linear in $n$) and required memory is $\mathcal{O}(nk)$ where $k$ is the number of neighbors.  Specifically, we order the $4314$ observation locations after which the joint distribution can be written as a sequential product form in conditional distributions.  For each location, the NNGP replaces the conditioning on all of the previous locations by conditioning only on the \emph{closest} $k$ previous locations.  Following empirical investigation in \cite{Dattaetal(16c)}, we set $k=15$ for the analysis below.
For sampling $\bm{\eta}$, we implement Metropolis-Hastings (MH) updates. The sampling distribution theory details for all parameters are described in the Supplementary Material (Appendix S4).
As for prior specifications, all are weak; we assume $\bm{\alpha}, \bm{\beta}\sim \mathcal{N}(\bm{0}, 100 \mathbf{I})$, $\delta\sim \mathcal{N}(\bm{0}, 100)$, $\sigma_{\omega}^2, \sigma_{\eta}^2\sim \mathcal{IG}(2,0.1)$ and $\phi_{\omega}, \phi_{\eta}\sim \mathcal{U}(0, 200)$. We set $\tau^2=1$ for the identifiability of the other parameters.
We discard the first 20,000 iterations as burn-in and preserve the subsequent 20,000 as posterior samples.
To provide adequate posterior inference, the MCMC iterations are tuned so that effective sample sizes (ESS) for all parameters are larger than 100. The ESS results reveal that the smallest value arises for the constant mean parameter $\alpha_{0}$, which is highly correlated with $\omega$ in models (b) and (c).

Table \ref{tab:delta} displays the estimation results for $\delta$ for models (c) and (d).
For MR, JB, GB and AO, the results for model (d) suggest significant preferential sampling effects; the means for $\delta$ are significantly different from $0$.
When $\delta > 0$, this implies that, in the selection  of the presence/absence locations for the species, presences were oversampled.  When $\delta < 0$, this means that in the selection  of the presence/absence locations for the species, presences were undersampled.
This insight is useful and can help in predicting probability of presence at unobserved locations. Furthermore, failing to include the $\eta(\bs)$ into the modeling might lead to misinterpretation of the effects of the regressors.

The $\eta(\bs)$ also provide improved prediction of presence/absence (see below).
However, with inclusion of the $\omega(\bs)$ surface (model (c)), the $\delta$ coefficients become insignificant.  The flexibility of the $\omega(\bs)$ surface seems to remove the benefit of using the $\eta(\bs)$ surface as a predictor.

Table \ref{tab:coef} displays the estimation results for models (a) and (d) for the vector of regression coefficients, $\bm{\alpha}$, for MR, JB, GB and AO (each has a significant $\delta$ with model (d)).
Introducing the $\eta(\bs)$ surface affects the estimation results for $\bm{\alpha}$.
For example,  the estimated $\alpha$ of meanTDQ for GB is significantly negative for model (a) but becomes insignificant for model (d).



\begin{figure}[ht]
  \begin{center}
   \includegraphics[width=15cm]{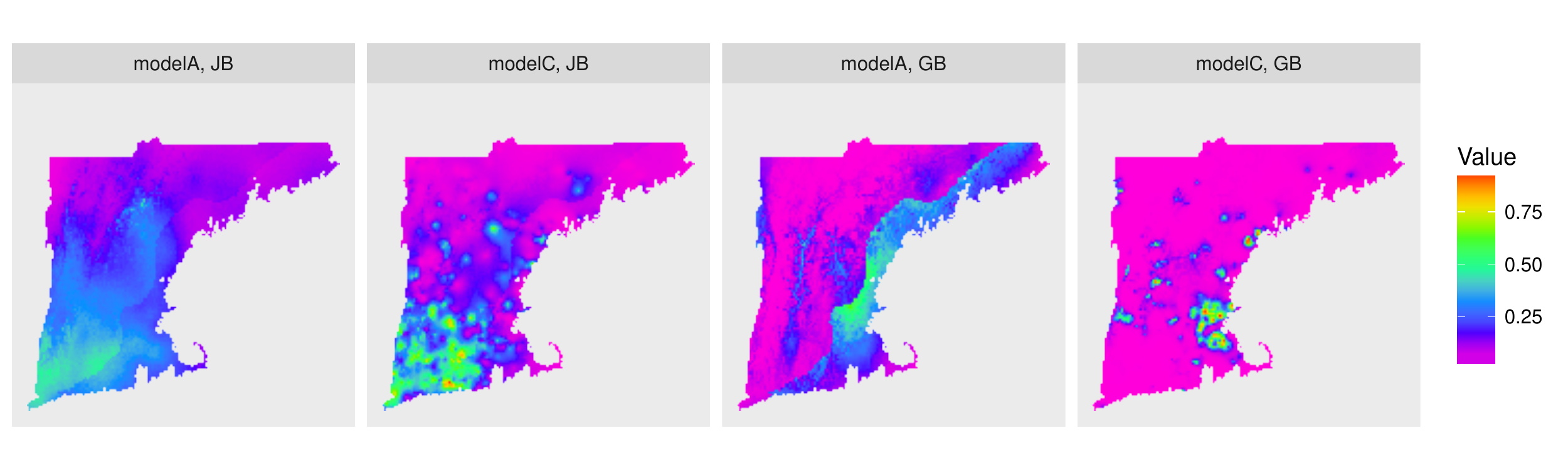}
  \end{center}
  \caption{}
  \label{fig:PAProb}
\end{figure}

Figure \ref{fig:PAProb} displays the posterior mean probability of presence surface under models (a) and (c) for JB and GB.
The surfaces for model (c) are very different from those for model (a), capturing local behavior.  That is, by comparison with Figure \ref{fig:PAlocations}, model (c) captures the presence probability better than model (a) which smooths away too much detail.  This point is supported through comparison of predictive performance below.  The model fitting results for the LGCP in (ii) are not of direct interest but, for completeness, are presented in the Supplementary Material (Appendix S1).


With binary response, to demonstrate improved prediction we consider misclassification error using the Tjur $R^2$ coefficient of determination \citep{Tjur(09)}.
This measure prefers a model with high probability of presence when presence is observed and low probability of presence when absence is observed.  For species $j$, this quantity is given by $TR_{j}=(\hat{\pi}_{j}(1)-\hat{\pi}_{j}(0))$ where $\hat{\pi}_{j}(1)$ and $\hat{\pi}_{j}(0)$ are the average probabilities of presence for the observed ones and zeros associated with the $j$-th species across the locations.
The larger the $TR_{j}$ , the better the discrimination.

We held out $20\%$ (879) of the presence/absence locations, chosen at random, for the seven species.
Table \ref{tab:pred} displays the results for the TR measure for models (a) - (d).
For all species, the models including $\omega(\bs)$, (b) and (c), outperform those without, (a) and (d).  Model (c) with (ii) tends to be better than model (b) (which ignores (ii)) particularly for AO, BB, and GM.  Model (d) with (ii) is at least as good as model (a) (which ignores (ii)) but is really only consequentially better for species GB which has the largest $\delta$ coefficient under model (d).



\section*{Fusing presence/absence and presence-only data}
\label{sec:fusion}

We turn to the data fusion question.  Data fusion (also \emph{assimilation}) is a widely employed objective when multiple data sources are available to inform about a common response of interest \citep[][]{NychkaAnderson(10), WikleBerliner(07)}.  A canonical example is the goal of modeling exposure to an environmental contaminant when we might have multiple data sources for the contaminant.  For instance, with ozone exposure, we may have data available from a network of monitoring stations, we may also have available output from a computer model built from theory of atmospheric transport, and we might have satellite data available from programs such as MODIS or Landsat.  The conceptual modeling strategy is to model a latent \emph{true} exposure surface and then build a model for each data source, conditioned upon the true surface.  The joint modeling enables each of the sources to inform about the true exposure surface, to enable improved prediction of this surface.  Examples in the literature include application to weather  data, sea surface temperature, and animal behavior patterns \citep[][]{Wikleetal(01), Sahuetal(10), Rundeletal(15)}.

In our setting, data fusion is different from customary settings.  Rather than multiple data sources informing about a common response, e.g., ozone level, we have two different types of data.  While both inform about species distribution, we have argued above that presence/absence data is not described stochastically in the same way as presence-only data.  The fusion approaches considered in the literature \citep[][]{Fithianetal(15), Dorazio(14), Giraudetal(16), Pacifici(17)} ignore this and assume a latent point pattern model for the presence-only data and that the presence/absence data is induced under this model.  Since we argue that a point pattern specification is inappropriate for presence/absence data, we think a different type of fusion is required.  We have a point pattern model for the presence-only data and a binary map model for the presence/absence data.  Since the point pattern of presences may inform about the probability of presence at a location, again we turn to preferential sampling ideas \citep[][]{Diggleetal(10)} in order to explore a coherent probabilistic fusion.

The extra information available to make a data fusion story is ${\cal S}_{PO}$, the set of observed presence-only locations.
Formally, what information does ${\cal S}_{PO}$ bring with regard to learning about the probability of presence surface?  As in ``What does ``probability of presence'' mean?'', assume that ${\cal S}_{PO} = \{\bs_{1}^{*},...,\bs_{m}^{*}\}$ is a complete census of species locations in $D$.  Associated with ${\cal S}_{PO}$,
we consider an intensity, $\lambda_{PO}(\bs)$, specified with a set of models similar to (i) or (ii) above.
We expect $\lambda_{PO}(\bs)$ to be elevated near these observations.
For example, analogous to (ii), let $\log \lambda_{PO}(\bs) = \bw^{T}(\bs)\bbeta_{PO} + \eta_{PO}(\bs)$, using the same predictors as with the presence/absence modeling.  However, the mechanisms that created ${\cal S}_{PO}$ and ${\cal S}_{PA}$ (the point pattern of presence/absence locations) are different, so it doesn't make sense that ${\cal S}_{PO}$ and ${\cal S}_{PA}$ follow the same model.  In order to capture the influence of ${\cal S}_{PO}$ on the $p(\bs)$ surface associated with ${\cal Y}_{PA}$ (the presence/absence data), we could add $\delta_{PO}\eta_{PO}(\bs)$ to the mean for $Z(\bs)$ in model (c) of ``Preferential sampling: Preferential sampling models for presence/absence data'', i.e. we could have a $\delta_{PA}\eta_{PA}(\bs)$ term and a $\delta_{PO}\eta_{PO}(\bs)$ term in order to improve prediction of presence/absence.

So, we have two sources for possible preferential sampling, one for each dataset.
We might insist that $\delta_{PO} >0$.  Then, from the presence-only data, the probability of presence will be increased around the $\bs_{j}^{*}$'s and decreased away from them.  Indeed, the locations in ${\cal S}_{PO}$ are severely biased; they are locations where we see only $1$'s.  We are severely over-sampling presences with ${\cal S}_{PO}$ and we should increase probability of presence where we do.

We adopt a model for ${\cal S}_{PO}$ analogous to model (ii) in ``Preferential sampling: Model fitting and inference for presence/absence data using preferential sampling'' for ${\cal S}_{PA}$
Then, we can add a $\delta_{PO}\eta_{PO}(\bs)$ term to the mean of $Z(\bs)$ under (b), (c), or (d).  In other words, we model ${\cal S}_{PA}$ as a LGCP with intensity
\begin{equation}
\log \lambda_{PA}(\bs) = \bw^{T}(\bs)\bbeta_{PA} + \eta_{PA}(\bs)
\label{eq:PA}
\end{equation}
and ${\cal S}_{PO}$ as a LGCP with intensity
\begin{equation}
\log \lambda_{PO}(\bs) = \bw^{T}(\bs)\bbeta_{PO} + \eta_{PO}(\bs).
\label{eq:PO}
\end{equation}

We consider the following models for ${\cal Y}$ arising directly through specification for $Z(\bs)$:\\

(c'): $Z(\bs) = \bx^{T}(\bs)\balpha + \delta_{PO} \eta_{PO}(\bs) + \omega(\bs) + \epsilon(\bs).$ \\

(d'): $Z(\bs) = \bx^{T}(\bs)\balpha + \delta_{PO} \eta_{PO}(\bs) + \epsilon(\bs).$ \\

These models replace $\delta_{PA}\eta_{PA}(\bs)$ with $\delta_{PO}\eta_{PO}(\bs)$ in models (c) and (d).  They allow only the point pattern of the presence-only data to help explain probability of presence.
In addition, we consider two models which also include preferential sampling associated with the presence/absence data, $\delta_{PA} \eta_{PA}(\bs)$: \\

(e): $Z(\bs) = \bx^{T}(\bs)\balpha + \delta_{PA} \eta_{PA}(\bs) + \delta_{PO} \eta_{PO}(\bs) + \epsilon(\bs)$ \\

(f): $Z(\bs) = \bx^{T}(\bs)\balpha + \delta_{PA} \eta_{PA}(\bs) + \delta_{PO} \eta_{PO}(\bs) + \omega(\bs) + \epsilon(\bs)$ \\
The Supplementary Material (Appendix S4) shows how to fit these models.

As a last remark, in practice, with a partial realization of the presence-only point pattern, we need to degrade $\lambda_{PO}(\bs)$ in the model fitting.  The following subsection briefly reviews an approach to implement such degradation.  The Supplementary Material (Appendix S4) shows how to adjust the fitting of  the models above in the presence of a partially observed presence-only point pattern.

\subsection*{Spatial modeling of presence-only data in practice}

Analysis of presence-only data has seen growth in recent years due to increased availability of such records from museum databases and other non-systematic surveys, see \cite{Grahametal(04)}.  Presence-only data is not \textit{inferior} to presence/absence data.  In fact, it can be viewed as the opposite; in principle, presence-only data offer a complete census while presence/absence data, since confined to a specified set of sampling sites, contains less information.  However, in practice, a complete census of individuals is rarely achieved.  The sampling effort required to obtain such censuses usually exceeds the available resources.

An early model-based strategy for presence-only data attempts to implement a presence/absence approach by drawing so-called \emph{background samples}, a random sample of locations in the region with known environmental features.  These samples were characterized as pseudo-absences \citep[][]{Engleretal(04), Ferrieretal(02)} and a logistic regression was fitted to the observed presences and these pseudo-absences, following ``Some presence/absence modeling details''.  Since presence/absence is unknown for these samples, work of \cite{PearceBoyce(06)} and \cite{Wardetal(09)} showed how to adjust the resulting logistic regression to account for this.  In any event, this approach \emph{manufactures} an arbitrary amount of data.  Additionally, it ignores spatial dependence for presence/absence across locations.  The observed presences, as a random number of random locations, should be viewed as a spatial point pattern \cite[see][in this regard]{WartonShepherd(10), Chakrabortyetal(11)}.

An algorithmic strategy in common use these days is the maximum entropy (Maxent) approach, \citep[see, e.g.,][]{Phillipsetal(06), Phillipsetal(09)}.  Maxent is a constrained optimization method which finds the optimal species density (closest to a uniform) subject to moment constraints.
The availability of an attractive software package (http://biodiversityinformatics.amnh.org /open$\_$source/maxent/), encourages its use for presence-only data analysis.  The resultant density surface is interpreted as providing the relative chance of observing a species at a given location compared to other locations in the region (and can not be interpreted as providing presence/absence probabilities).  However, as an optimization strategy rather than a stochastic modeling approach, Maxent is unable to attach any uncertainty to resulting optimized estimates.  Also, Maxent is unable to provide an intensity surface. Hence, for example, we are unable to determine the expected number of individuals in a specified region or the probability of at least one individual in a specified region.

Arguably, a formal point pattern modeling approach is preferable since it enables full inference, with associated uncertainty, over the region.
Modeling presence-only data as a point pattern specifies an associated intensity in terms of the available environments, at available spatial scale, across the region.  Spatial structure for the intensity surface is introduced through spatial random effects, resulting in a log Gaussian Cox process \citep[][]{Molleretal(98), MollerWaagepetersen(04)}, as discussed in ``Preferential sampling'' above.

Employing the LGCP in practice acknowledges that the observed point pattern is biased through anthropogenic processes, e.g., human intervention to transform the landscape and non-uniform (in fact, often very irregular) sampling effort.  Such bias in sampling is a common problem, see for example \cite{Loiselleetal(08)} and references therein. This requires adjusting the \emph{potential} species intensity to a \emph{realized} intensity which is treated as a \emph{degradation} of the former.  Such modeling adjustment is discussed in detail in \cite{Chakrabortyetal(11)} which we briefly review below.

Variation in site access is one of the factors that influences the likelihood of the site to be visited/sampled.  For example, sites (i) adjacent to roads or along paths, (ii) near urban areas, (iii) with public ownership, e.g., state or national parks, or (iv) with flat topography are likely to be over-sampled relative to more inaccessible sites.  When bias implies that only a portion of the region is sampled, it is likely that only a portion of the overall point pattern is observed.
Land use, as a result of human intervention, affects \emph{availability} of locations, hence, also inference about the intensity.  Also, agricultural transformation and dense stands of alien invasive species preclude availability.  Transformed areas are not sampled and this information must also be included in the modeling.  Altogether, sampling tends to be sparse and irregular; we rarely collect a random sample of all available environments.

Detection can affect inference regarding the intensity.  That is, we may incorrectly identify a species as present when it is actually absent (false presence) or fail to detect a species that is actually present (false absence) \citep[][]{Reeseetal(05)}.  The prevalence of these false records will affect the performance of an explanatory model on response to environmental features \citep[][]{Tyreetal(03)}. Modeling for these errors can be attempted but requires information beyond the scope here.

\subsection*{Some explicit modeling details}
\label{sec:effort}
Following ideas in \cite{Chakrabortyetal(11)}, we conceptualize a \emph{potential} intensity, i.e., the intensity in the absence of degradation, as well as a \emph{realized} (or effective) intensity that operates in the presence of degradation.  The intensity is tiled to grid cells at the resolution of the available environmental covariate surface.
We consider three surfaces over a region of interest, $D$.  First, let $\lambda_{PO} (\bs)$ be the potential intensity surface, i.e., a positive function which is integrable over $D$.  $\lambda_{PO}(\bs)$ is the intensity in the absence of degradation.  With  $\int_{D} \lambda_{PO}(\bs) d\bs = \lambda_{PO}(D)$, $g(\bs) = \lambda_{PO}(s)/\lambda_{PO}(D)$ gives the potential density over $D$. Modeling for $\lambda_{PO}(\bs)$ is a log Gaussian Cox process (LGCP) which incorporates environmental covariates, $\mathbf{x}(\bs)$, to influence the intensity as a linear form in parameters. So, for any location $s \in D$, as in ``Preferential sampling'' above, we have
$\text{log}\lambda_{PO}(\bs) = \mathbf{x}^{T}(\bs)\bbeta + \omega(\bs)$
with $\omega(\bs)$, a zero-mean stationary, isotropic Gaussian process (GP) over $D$, to capture residual spatial association in the $\lambda_{PO}(\bs)$ surface across grid cells.

Turning to degradation, we envision an availability surface, $U(\bs)$, a binary surface over $D$ such that $U(\bs) =1$ or $0$ according to whether location $\bs$ is untransformed (hence, available) by land use or not.  That is, assuming no sampling bias, $\lambda_{PO}(\bs) U(\bs)$ can only be $\lambda_{PO}(\bs)$ or $0$ according whether $\bs$ is available or not.  Thirdly, we envision a sampling effort surface over $D$ which we denote as $T(\bs)$.  $T(\bs)$ is also a binary surface and $T(\bs)U(\bs)=1$ indicates that location $\bs$ is both available and sampled.  Altogether, $\lambda_{PO}(\bs) U(\bs) T(\bs)$ becomes the degraded intensity at location $\bs$. This implies that in regions where no locations were sampled, the degraded intensity for the species is $0$.

We partition $D$ into grid cells with $A_i, i=1,2,...I$ denoting the geographical region corresponding to grid cell $i$.  Typically the gridding is at the resolution of the predictors used in explaining $\lambda_{PO}(\bs)$.  Then, if we average $U(\bs)$ over $A_i$, we obtain $u_{i} = \int_{A_{i}} U(\bs)d\bs/|A_{i}|$ where $|A_i|$ is the area of cell $i$.  So,  $u_{i}$ is the proportion of cell $i$ that is transformed. The $u_i$ can often be obtained, through remote sensing.   Further, we can set $q_i = \int_{A_{i}} T(\bs)U(\bs) d\bs/|A_{i}|$ and interpret $q_i$ as the probability that a randomly selected location in $A_i$ was available and sampled.  Thus, we can capture availability and sampling effort at areal unit scale.
Additionally, $\int_{A_{i}} T(\bs)d\bs/|A_{i}|$ can be viewed as the sampling probability associated with cell $i$.  Then, if $T(\bs)$ is viewed as random, the expectation of the integral would yield $\int_{A_{i}} p(\bs) d\bs/|A_i|$ where, now, $p(\bs) = P(T(\bs)=1) \in [0,1]$.  $p(\bs)$ gives the local probabilities of sampling, not a probability density over $D$.
Finally, if we define $p_{i}$ through $q_{i} = u_{i}p_{i}$, then $p_{i} = \frac{\int_{A_{i}} T(\bs)U(\bs) d\bs}{\int_{A_{i}}U(\bs)d\bs}$, i.e., $p_i$ is the conditional probability that a randomly selected location in cell $i$ is sampled given it is available.  As an illustration, we might set $p_i$ equal to $1$ or $0$ which we interpret as $T(\bs)=U(\bs), \: \forall \bs \in A_{i}$ or $T(\bs)=0, \: \forall \bs \in A_{i}$, respectively.  That is, either all available sites in $A_i$ were visited or no available sites in $A_i$ were visited.  This degraded point pattern model is what we use for the data fusion.  Fitting is described briefly in the Supplementary Material (Appendix S4).  Full details are provided in \cite{Chakrabortyetal(11)}

%

\subsection*{Model fitting and inference for data fusion}
\label{sec:fitting-fusion}
We fit the models above in \eqref{eq:PA} and \eqref{eq:PO} with models (a), (b), (c'), (d'), (e), and (f) with details given in the Supplementary Materials Appendix S4.  We implement the degradation as discussed in ``Fusing presence/absence and presence-only data: Some explicit modeling details'' for the presence-only data.  In the absence of information about land use, we assume $U(s)=1$ for all locations.
There is no simple solution for modeling sampling effort.  However, in the absence of a complete census, some assumption needs to be made in order to sensibly degrade the intensity.  We adopt the sampling effort surface $T(\bs)$ for each grid cell such that $T(\bs)=1$ for all cells where at least one presence-only point is observed across all species, $T(\bs)=0$ otherwise.   There is nothing in the modeling that, for a given species, prevents/discourages $\lambda_{PO}(\bs)$ from being small if the data suggests it.  We are only attempting to account for degradation apart from this.

The estimation and predictive performance results for models (a) and (b) are the same as those in ``Preferential sampling: Model fitting and inference for presence/absence data using preferential sampling''.
Since $\delta_{PO}$ is expected to be positive, a priori, we adopt a truncated normal prior on the nonnegative domain, i.e., $\delta_{PO} \sim \mathcal{N}_{\ge 0}(0, 100)$.

Table \ref{tab:coef-fusion} displays the estimation results for model (e) which include both $\delta_{PA}\eta_{PA}(\bs)$ and $\delta_{PO}\eta_{PO}(\bs)$.   None of the $\delta_{PA}$ are significantly different from $0$.  However, all of the $\delta_{PO}$ are significantly different from $0$, revealing that the locations of the presence-only sites significantly improve the performance of the presence-absence model.
Table \ref{tab:pred-fusion} displays the results for the TR measure under the same settings as in ``Preferential sampling: Model fitting and inference for presence/absence data using preferential sampling''.
The results are similar to those in Table \ref{tab:pred}.  Performance is essentially indistinguishable across all models other than model (a); however, model (f) emerges as the best.  As a last remark here, if we focus on presence/absence locations which are near observed presence-only locations, we find an improvement in the TR measure for presence-absence at those locations compared to the corresponding model ignoring the presence-only data (results not shown).

\section*{Summary and future work}


Our contribution is to attempt to bring more clarity to a frequent activity for ecologists, modeling presence/absence for species, confining ourselves to plants.  We have done this from a probabilistic perspective, arguing that presence/absence data should be viewed as a point level phenomenon and therefore, stochastic modeling for presence/absence should be done at dimensionless points.  In the development we have also argued that attempting to model presence/absence at areal scale raises challenges and, further, that any such modeling is incompatible with point-level modeling. We have also asserted that the number of presences in a fixed bounded region must be finite and therefore, that a physical realization of a presence in the region is larger than a dimensionless point.

We acknowledge that we are being more formal in developing this perspective than is customary.  When presence/absence data is supplied at point-level, it will be a geo-coded location and, in many cases, it is supplied at areal scale, recording presence of the species anywhere in the areal unit.  However, this is not a deterrent from considering our perspective.  All continuous measurements are obtained up to rounding error.  When a temperature is recorded at a location, the location is provided up to the accuracy of the geo-coding device; nonetheless, we routinely model temperatures at (dimensionless) points.

Next, we turned our attention to attempting to improve prediction of probability of presence at a location for a presence/absence dataset.  We introduced the usefulness of preferential sampling in this context, anticipating that there may be bias in sampling sites visited for presence/absence data; sampling may favor seeing more presences.  We argued that the idea of a shared process model, viewing the set of presence/absence locations as a point pattern, can improve inference regarding the presence/absence surface.  We demonstrated this with a plant presence/absence dataset from New England.

Finally, we asserted that presence-only data should be modeled as a point pattern, albeit degraded due to availability and sampling effort over the study region.  We showed that, as a result, a common model for presence/absence and for presence-only data can not be stochastically coherent.  Hence, if we seek a data fusion having both presence/absence data as well as presence-only data, a different approach is needed.  We argued that, again, a shared process specification is coherent for such fusion and illustrated by adding a presence-only plant dataset from New England to the presence/absence dataset.

Future work offers much opportunity.  More experience is needed with regard to the rich set of modeling specifications that we have presented in Sections ``Preferential sampling'', ``Fusing presence/absence and presence-only data'' and Supplementary Material Appendix S2.  We also anticipate the need to supply user-friendly software to enable ecologists to play with these models with their own datasets.  A particularly useful future direction leads us to joint species distribution models.  These are easy to envision but challenging to fit.  Another useful future direction will consider different types of response data, e.g., abundance or basal area, where preferential sampling of locations may occur.  Possibly the most difficult challenge will be to move to animal movement data where the concepts of occupancy, use, and dynamics need to be carefully brought into play.


\section*{Acknowledgement}

The authors thank John Silander for numerous useful conversations which motivated much of the formalism presented here.  They also thank Jenica Allen  for supplying the IPANE and GBIF datasets which illustrated our ideas. The computational results were obtained using Ox version 6.21 \citep{Doornik(07)}.

\clearpage
\bibliographystyle{chicago}
\bibliography{SP}

\clearpage
\section*{Tables}
\begin{table}[ht]
\caption{Study species and sample sizes}
\label{tab:name}
\centering
\begin{tabular}{lcccccc}
\hline
\hline
Common & symbol & IPANE & IPANE & GBIF \\
name  & & presences & absences & presences \\
\hline
multiflora rose & MR & 1230 & 3084 & 249 \\
oriental bittersweet & OB & 1106 & 3208 & 305 \\
Japanese barberry & JB & 1012 & 3302 & 399 \\
glossy buckthorn & GB & 755 & 3559 & 223 \\
autumn olive & AO & 386 & 3928 & 193 \\
burning bush & BB & 336 & 3978 & 257 \\
garlic mustard & GM & 279 & 4035 & 440 \\
\hline
\hline
\end{tabular}
\end{table}


\clearpage
\begin{table}[htbp]
\caption{Models}
\small
\centering
\begin{tabular}{ll}
\hline
\hline
  Modeling for $\mathcal{S}$  & Modeling for $\mathcal{Y}$ \\
\hline
 (i) $\log \lambda(\bs)=\bw^{T}(\bs)\bbeta$ & (a) $Z(\bs) = \bx^{T}(\bs)\balpha + \epsilon(\bs)$ \\
 (ii) $\log \lambda(\bs)=\bw^{T}(\bs)\bbeta+\eta(\bs)$ & (b) $Z(\bs) = \bx^{T}(\bs)\balpha + \omega(\bs) + \epsilon(\bs)$ \\
 (iii) $\log \lambda(\bs) = \bw^{T}(\bs)\bbeta + \psi \omega(\bs)$ & (c) $Z(\bs) = \bx^{T}(\bs)\balpha + \delta \eta(\bs) +  \omega(\bs) + \epsilon(\bs)$ \\
 (iv) $\log \lambda(\bs)= \bw^{T}(\bs)\bbeta + \eta(\bs) + \xi(\bs)$ & (d) $Z(\bs) = \bx^{T}(\bs)\balpha + \delta \eta(\bs) + \epsilon(\bs)$ \\
\hline
\hline
\end{tabular}
\label{tab:Models}
\end{table}

\clearpage
\begin{table}[ht]
\caption{Estimation results for $\delta$ with models (c) and (d)}
\label{tab:delta}
\centering
\begin{tabular}{lccccccc}
\hline
\hline
  & \multicolumn{2}{c}{Model (c)}  & & \multicolumn{2}{c}{Model (d)}  \\
Species   & Mean & $95\%$ Int & & Mean  & $95\%$ Int  \\
\hline
MR & 0.028  & [-0.029, 0.092] & & 0.037 & \textbf{[0.023, 0.071]}  \\
OB & -0.014  & [-0.072, 0.044] & & -0.027  & [-0.063, 0.006]  \\
JB & 0.024  & [-0.043, 0.093] && 0.085  & \textbf{[0.048, 0.122]}  \\
GB & 0.075  & [-0.052, 0.194] && 0.225 & \textbf{[0.179, 0.274]}  \\
AO & -0.049  & [-0.133, 0.039] && -0.076 & \textbf{[-0.120, -0.030]}  \\
BB & 0.064  & [-0.025, 0.164] && 0.013  & [-0.033, 0.062] \\
GM & 0.041  & [-0.100, 0.213] && -0.036  & [-0.085, 0.015]  \\
\hline
\hline
\end{tabular}
\end{table}

\clearpage
\begin{table}[ht]
\caption{Estimation results for MR, JB, GB, and AO for models (a) and (d). The bold font suggests the change of significance.}
\label{tab:coef}
\centering
\scalebox{0.8}{
\begin{tabular}{lcccccccc}
\hline
\hline
   & \multicolumn{2}{c}{MR} & \multicolumn{2}{c}{JB} & \multicolumn{2}{c}{GB} & \multicolumn{2}{c}{AO}  \\
Model(a)     & Mean  & $95\%$ Int & Mean  & $95\%$ Int  & Mean  & $95\%$ Int & Mean  & $95\%$ Int \\
\hline
const & -1.774 & [-1.951, -1.602] & -1.507  & [-1.676, -1.344] & -1.906 & [-2.126, -1.693] & -2.360 & [-2.634, -2.102]  \\
mDR & 0.193 & [0.027, 0.362] & 0.383 & [0.213, 0.556] & 0.110 & [-0.082, 0.305] & 0.381 & [0.161, 0.603]   \\
maxTWM & -0.020 & [-0.217, 0.175] & -0.236  & [-0.435, -0.035] & 0.463 & [0.223, 0.704] & -0.554 & [-0.813, -0.300]   \\
meanTDQ & 0.715 & [0.460, 0.970] &0.762  & [0.498, 1.028] & -0.324 & [-0.619, -0.025] & 0.859 & [0.532, 1.196]  \\
minTCM & -0.070 & [-0.149, 0.010] & -0.065  & [-0.146, 0.015] & 0.415 & [0.309, 0.522] & 0.029 & [-0.074, 0.136]  \\
PWM & 0.178 & [0.092, 0.262] & 0.063  & [-0.024, 0.151] & -0.139 & [-0.240, -0.038] & -0.182 & [-0.300, -0.066]  \\
PS & -0.142 & [-0.246, -0.040] & -0.083  & [-0.182, 0.017] & -0.030 & [-0.141, 0.081] & -0.162 & [-0.320, -0.010]  \\
PWQ & -0.062 & [-0.166, 0.042] & 0.204  & [0.103, 0.307] & -0.371 & [-0.502, -0.239] & -0.156 & [-0.308, -0.001]  \\
\hline
Model(d)     & Mean  & $95\%$ Int & Mean  & $95\%$ Int  & Mean  & $95\%$ Int & Mean  & $95\%$ Int \\
\hline
const & -1.931 & [-2.156, -1.705] & -1.845  & [-2.072, -1.620] & -2.985 & [-3.331, -2.654] & -2.066 & [-2.391, -1.751]  \\
mDR & 0.215 & [0.043, 0.386] &0.423  & [0.246, 0.597] & \textbf{0.352} & \textbf{[0.124, 0.578]} & 0.362 & [0.141, 0.585]   \\
maxTWM & -0.049 & [-0.247, 0.145] & -0.308  & [-0.514, -0.102] & \textbf{0.149} & \textbf{[-0.135, 0.433]} & -0.507 & [-0.760, -0.252]   \\
meanTDQ & 0.792 & [0.524, 1.054] &0.900  & [0.622, 1.174] & \textbf{0.206} & \textbf{[-0.145, 0.550]} & 0.756 & [0.418, 1.089]  \\
minTCM & -0.063 & [-0.146, 0.021] & -0.051  & [-0.137, 0.034] & 0.480 & [0.350, 0.612] & 0.020 & [-0.087, 0.129]  \\
PWM & 0.161 & [0.073, 0.250] & 0.022  & [-0.070, 0.114] & -0.225 & [-0.371, -0.096] & -0.148 & [-0.269, -0.026]  \\
PS & -0.132 & [-0.238, -0.026] & -0.061  & [-0.165, 0.040] & 0.046 & [-0.083, 0.183] & -0.174 & [-0.333, -0.017]  \\
PWQ & -0.028 & [-0.141, 0.083] &0.267  & [0.153, 0.381] & -0.218 & [-0.399, -0.032] & -0.195 & [-0.357, -0.038]  \\
$\delta$ & 0.037 & [0.023, 0.071] & 0.085  & [0.048, 0.122] & 0.225 & [0.179, 0.274] & -0.076 & [-0.120, -0.030]  \\
\hline
\hline
\end{tabular}}
\end{table}

\clearpage
\begin{table}[ht]
\caption{Estimation results for the TR measure for preferential sampling.}
\label{tab:pred}
\centering
\scalebox{0.9}{
\begin{tabular}{lcccccccccc}
\hline
\hline
  & \multicolumn{2}{c}{Model (a)} & \multicolumn{2}{c}{Model (b)} & \multicolumn{2}{c}{Model (c)}  & \multicolumn{2}{c}{Model (d)}  \\
Species     & Mean  & $95\%$ Int  & Mean  & $95\%$ Int & Mean  & $95\%$ Int & Mean  & $95\%$ Int  \\
\hline
MR & 0.104  & [0.094, 0.114] & 0.168 & [0.145, 0.191] & \textbf{0.176} & [0.157, 0.202]  & 0.105  & [0.096, 0.114] \\
OB & 0.099  & [0.088, 0.109] & \textbf{0.183} & [0.163, 0.200] & \textbf{0.183} & [0.158, 0.203]  & 0.101  & [0.092, 0.111] \\
JB & 0.072  & [0.063, 0.081] & \textbf{0.201} & [0.180, 0.230] & 0.198 & [0.169, 0.227]  & 0.075  & [0.066, 0.083] \\
GB & 0.126  & [0.112, 0.139] & 0.405 & [0.372, 0.434] & \textbf{0.412} & [0.382, 0.451] & 0.162 & [0.147, 0.175] \\
AO & 0.034  & [0.027, 0.043] & 0.095 & [0.068, 0.129] & \textbf{0.115} & [0.073, 0.156] & 0.039  & [0.033, 0.051] \\
BB & 0.057  & [0.048, 0.068] & 0.112 & [0.078, 0.152] & \textbf{0.131} & [0.097, 0.168] & 0.057  & [0.049, 0.066] \\
GM & 0.026  & [0.020, 0.032] & 0.135 & [0.111, 0.166] & \textbf{0.164} & [0.118, 0.206] & 0.027 & [0.022, 0.033] \\
\hline
\hline
\end{tabular}}
\end{table}

\clearpage
\begin{table}[ht]
\caption{Estimation results for data fusion model (e).}
\label{tab:coef-fusion}
\centering
\scalebox{0.80}{
\begin{tabular}{lcccccccccc}
\hline
\hline
   & \multicolumn{2}{c}{MR} & \multicolumn{2}{c}{OB} & \multicolumn{2}{c}{JB} & \multicolumn{2}{c}{GB} \\
Model(e)     & Mean  & $95\%$ Int  & Mean  & $95\%$ Int & Mean  & $95\%$ Int & Mean  & $95\%$ Int \\
\hline
const & -2.282  & [-2.670, -1.894] & -2.148 & [-2.560, -1.724] & -1.867 & [-2.298, -1.479] & -3.112 & [-4.153, -2.226] \\
mDR & 0.134  & [-0.160, 0.419] & 0.264 & [-0.088, 0.607] & 0.313 & [-0.036, 0.644]  & 0.502 & [-0.196, 1.252] \\
maxTWM & 0.147  & [-0.185, 0.490] & -0.012 & [-0.419, 0.410] & -0.030 & [-0.431, 0.398] & -0.378 & [-1.310, 0.635]  \\
meanTDQ & 0.745  & [0.280, 1.195] & 0.788 & [0.235, 1.343] & 0.659 & [0.112, 1.209] & 0.174 & [-0.930, 1.296] \\
minTCM & -0.023  & [-0.165, 0.123] & 0.052 & [-0.122, 0.230] & -0.066 & [-0.234, 0.094] & 0.525 & [0.107, 0.948] \\
PWM & 0.098  & [-0.046, 0.247] & 0.067 & [-0.108, 0.248] & 0.060 & [-0.116, 0.236] & -0.399 & [-0.881, 0.037] \\
PS & -0.207  & [-0.370, -0.042] & -0.077 & [-0.281, 0.123] & -0.128 & [-0.316, -0.060] & 0.033 & [-0.385, 0.442] \\
PWQ & -0.006  & [-0.179, 0.170] & -0.005 & [-0.228, 0.212] & 0.252 & [0.053, 0.453] & -0.041 & [-0.523, 0,438] \\
$\delta_{PA}$ & 0.034  & [-0.025, 0.093] & -0.030 & [-0.093, 0.027] & -0.003 & [-0.072, 0.063] & 0.107 & [-0.000, 0.231] \\
$\delta_{PO}$ & 0.463  & [0.362, 0.582] & 0.957 & [0.652, 1.449] & 0.831 & [0.654, 1.029] & 1.350 & [1.042, 1.905] \\
\hline
   & \multicolumn{2}{c}{AO} & \multicolumn{2}{c}{BB} & \multicolumn{2}{c}{GM} & \multicolumn{2}{c}{} \\
Model(e)     & Mean  & $95\%$ Int  & Mean  & $95\%$ Int & Mean  & $95\%$ Int &  &  \\
\hline
const & -2.904  & [-3.632, -2.310] & -3.322 & [-4.061, -2.734] & -3.774 & [-4.977, -2.887] &  &  \\
mDR & 0.373  & [-0.052, 0.776] & -0.087 & [-0.474, 0.301] & 0.348 & [-0.197, 0.927]  &  &  \\
maxTWM & -0.381 & [-0.866, 0.146] & 0.740 & [0.245, 1.259] & 0.061 & [-0.608, 0.726] &  &   \\
meanTDQ & 0.715  & [0.050, 1.343] & 0.158 & [-0.451, 0.790] & 1.065 & [0.127, 2.037] &  &  \\
minTCM & 0.032  & [-0.173, 0.229] & 0.050 & [-0.139, 0.238] & -0.024 & [-0.297, 0.253] &  &  \\
PWM & -0.127  & [-0.339, 0.087] & 0.128 & [-0.092, 0.356] & -0.393 & [-0.737, -0.062] &  & \\
PS & -0.320  & [-0.577, -0.057] & -0.343 & [-0.603, -0.090] & -0.063 & [-0.397, 0.271] & & \\
PWQ & -0.368  & [-0.663, -0.093] & 0.270 & [-0.015, 0.564] & 0.601 & [0.225, 1.050] &  &  \\
$\delta_{PA}$ & -0.039  & [-0.118, 0.040] & 0.035 & [-0.043, 0.124] & 0.019 & [-0.098, 0.161] &  &  \\
$\delta_{PO}$ & 0.703  & [0.508, 0.933] & 0.817 & [0.587, 1.121] & 1.053 & [0.841, 1.342] &  &  \\
\hline
\hline
\end{tabular}}
\end{table}

\clearpage
\begin{table}[ht]
\caption{Estimation results for the TR measure for the data fusion. The results for (a) and (b) are the same as in Table \ref{tab:pred}}
\label{tab:pred-fusion}
\centering
\scalebox{0.9}{
\begin{tabular}{lcccccccccc}
\hline
\hline
   & \multicolumn{2}{c}{Model (a)} & \multicolumn{2}{c}{Model (b)} & \multicolumn{2}{c}{Model (c)}  \\
Species     & Mean  & $95\%$ Int & Mean  & $95\%$ Int  & Mean  & $95\%$ Int  \\
\hline
MR & 0.104 & [0.094, 0.114] & 0.168 & [0.145, 0.191] & 0.171 & [0.154, 0.193]   \\
OB & 0.099 & [0.088, 0.109] & \textbf{0.183} & [0.163, 0.200] & 0.178 & [0.155, 0.198]   \\
JB & 0.072 & [0.063, 0.081] & \textbf{0.201} & [0.180, 0.230] & 0.198 & [0.175, 0.223]  \\
GB & 0.126 & [0.112, 0.139] & 0.405 & [0.372, 0.434] & 0.407 & [0.372, 0.436]  \\
AO & 0.034 & [0.027, 0.043] & 0.095 & [0.068, 0.129] & 0.103 & [0.075, 0.139]  \\
BB & 0.057 & [0.048, 0.068] & 0.112 & [0.078, 0.152] & 0.127 & [0.099, 0.163]   \\
GM & 0.026 & [0.020, 0.032] & 0.135 & [0.111, 0.166] & 0.143 & [0.112, 0.182]  \\
\hline
   & \multicolumn{2}{c}{Model (d)} & \multicolumn{2}{c}{Model (e)} & \multicolumn{2}{c}{Model (f)}   \\
     & Mean  & $95\%$ Int & Mean  & $95\%$ Int  & Mean  & $95\%$ Int  \\
\hline
MR  & 0.170 & [0.147, 0.195] & 0.165  & [0.136, 0.195] & \textbf{0.171} & [0.152, 0.197]  \\
OB  & 0.172 & [0.151, 0.197] & 0.169  & [0.134, 0.190] & 0.179 & [0.152, 0.207]  \\
JB  & 0.188 & [0.165, 0.216] & 0.186  & [0.159, 0.213] & \textbf{0.201} & [0.172, 0.221]  \\
GB  & 0.401 & [0.366, 0.440] & 0.404  & [0.363, 0.441] & \textbf{0.412} & [0.372, 0.449]  \\
AO  & 0.097 & [0.067, 0.137] & 0.101  & [0.073, 0.133] & \textbf{0.106} & [0.071, 0.144] \\
BB  & 0.124 & [0.097, 0.158] & 0.124  & [0.085, 0.169] & \textbf{0.134} & [0.093, 0.169]  \\
GM  & 0.144 & [0.101, 0.189] & 0.150  & [0.102, 0.195] & \textbf{0.156} & [0.115, 0.197] \\
\hline
\hline
\end{tabular}}
\end{table}

\clearpage
\section*{Figures}
Figure captions
\begin{itemize}
\item Figure 1: The distribution of presence (blue) and absence (green) points for each species across the study region. 
\item Figure 2: The distribution of presence only points (red) for each species across the study region. 
\item Figure 3: The standardized covariate surface for each of the 7 selected covariates across the study region. 
\item Figure 4: Examples of sample locations and the latent surface $Z(\cdot)$: completely random sample (left), clustering (middle) and preferential (right).
For preferential sampling, $\log \lambda(\bs)=3+\eta(\bs)$ where $\eta(\bs)$ is simulated as zero mean Gaussian process realization with exponential covariance function, $C(\bm{s},\bm{s}')=3\exp(-\|\bm{s}-\bm{s}'\|)$.
The latent surface is specified as $Z(\bs)=\eta(\bs)$, i.e., shared process with $\delta=1$.
\item Figure 5: The posterior mean presence probability surface for models (a) and (c) for JB and GB
\end{itemize}

\clearpage

\clearpage

\clearpage
\end{document}